# SCIENTIFIC DATA

OPEN : Data Descriptor: **A multilayer network dataset of interaction and influence spreading in a virtual world**

Jarosław Jankowski[1], Radosław Michalski[2] & Piotr Bródka[2]

Presented data contains the record of five spreading campaigns that occurred in a virtual world platform. Users distributed avatars between each other during the campaigns. The processes varied in time and range and were either incentivized or not incentivized. Campaign data is accompanied by events. The data can be used to build a multilayer network to place the campaigns in a wider context. To the best of the authors' knowledge, the study is the first publicly available dataset containing a complete real multilayer social network together, along with five complete spreading processes in it.

| Design Type(s) | innate behavior design • decision analysis study design |
|---|---|
| Measurement Type(s) | Influence |
| Technology Type(s) | data collection method |
| Factor Type(s) | Incentive |
| Sample Characteristic(s) | Homo sapiens • Virtual Reality |

[1]West Pomeranian University of Technology, Department of Computer Science and Information Technology, Szczecin 71-210, Poland. [2]Wrocław University of Science and Technology, Department of Computational Intelligence, Wrocław 50-370, Poland. Correspondence and requests for materials should be addressed to J.J. (email: jjankowski@wi.zut.edu.pl).

## Background & Summary

Information spreading within electronic systems has attracted substantial research interest in recent years. Studies have investigated spreading processes in email exchange[1], social platforms[2], virtual worlds[3], online games[4], blogs and other forms of online publishing and communication[5]. The types of content include videos[6], images[7], news[8], information about promotions[9], rumours[10], and virtual goods[11–13]. Research in the field includes seeding strategies for selection of initial nodes which start diffusion processes[13,14], factors affecting the forwarding of content[15], the role of emotions[5] and the performance of viral campaigns[16].

Virtual worlds and online games create interesting environments having the ability to monitor content diffusion with mechanics close to the real systems. Users, represented by avatars in synchronic environments, perform actions similar to real world face-to-face communication, with direct contact between content sender and recipient. The level of interaction and the content sharing is influenced by factors similar to those occurring in the real world, such as visual characteristics, social ties, activity within groups and emotions. Virtual worlds can be treated as laboratories for analysis and observing social and economic phenomena[17]. Virtual worlds have been used for studying the dynamics of infectious diseases[18], virtual goods transmission[11,12], visual factors affecting the spread of information[3]. While several online repositories publish data from social networking platforms, datasets from virtual worlds (especially those related to information spread) are difficult to acquire. Virtual worlds and games have typically closed structures and lack open APIs to access data. One of the reasons is the fact that social networking platforms are dominated by leaders like Facebook or Twitter. In contrast, the online game market is fragmented with hundreds of systems which are available online[19]. Operators of online games and virtual worlds are often apathetic towards sharing internal data, thus making it difficult to do research in the field. Datasets presented in the current study reveal aspects of digital content spreading within virtual worlds showing several different mechanics. Data from viral campaigns, system use statistics and the multilayer structure of the network based on social relations, messages, virtual currency transfers and social interactions within virtual rooms each offer areas for research in the field of information spreading.

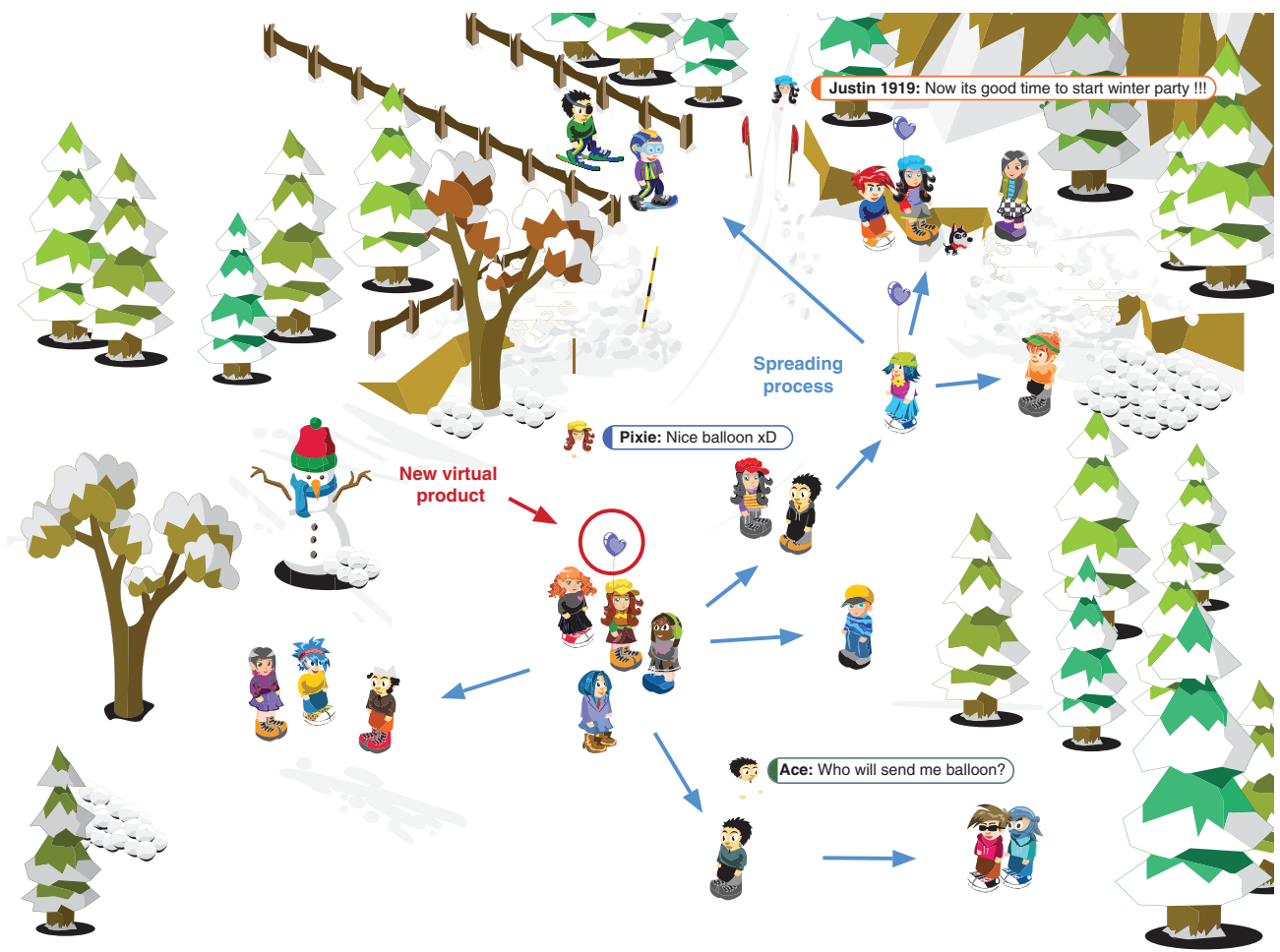

**Figure 1. Screenshot from virtual world with illustrated process of spreading of virtual product among users.**

| Id | Content type | Incentives | Quality | Mechanics | Duration |
|---|---|---|---|---|---|
| C1 | Occasional avatar related to the special Halloween event | No | Premium | Low resistance | 11 days |
| C2 | Occasional avatar related to the special Halloween event | Yes | Premium | Low resistance | 11 days |
| C3 | Winter avatar adjusted to winter theme and seasonal room | No | Basic | High resistance | 209 days |
| C4 | Thematic avatar for special anniversary event | No | Premium | High resistance | 204 days |
| C5 | User designed avatar with emotional appeal in a form of Guy Fawkes mask | No | Premium resistance | High | 154 days |

Table 1. Characteristics of campaigns.

| File | Lines | Size | Short description |
|---|---|---|---|
| campaigns | 4,981 | 189 KB | Semicolon delimited file containing five spreading processes in Timik.pl |
| friends | 12,416,809 | 474 MB | Semicolon delimited file containing the friendship relation between Timik.pl users |
| logins | 14,688,527 | 455 MB | Semicolon delimited file containing date and time of each user login to the service |
| messages | 26,134,695 | 999 MB | Semicolon delimited file containing the messages (without text) between Timik.pl users |
| transactions | 538,597 | 21.7 MB | Semicolon delimited file containing the virtual currency transactions between Timik.pl users |
| visits | 12,660,735 | 479 MB | Semicolon delimited file containing the visits of Timik.pl in private rooms of other users |

Table 2. Dataset dimensions.

| Column | Format | Short description |
|---|---|---|
| 1 | Integer | Id of the campaign (from 1 to 5) |
| 2 | DateTime | Date and time of Receiver activation |
| 3 | Integer | Activating user id in the Timik.pl service |
| 4 | Integer | Activated user id in the Timik.pl service |

Table 3. Description of campaigns file.

| Column | Format | Short description |
|---|---|---|
| 1 | DateTime | Date and time of establishing the friendship relation |
| 2 | Integer | Inviting user id in the Timik.pl service |
| 3 | Integer | Invited user id in the Timik.pl service |

Table 4. Description of friends file.

| Column | Format | Short description |
|---|---|---|
| 1 | DateTime | Date and time of login to the Timik.pl service |
| 2 | Integer | User id in the Timik.pl service |

Table 5. Description of logins file.

| Column | Format | Short description |
|---|---|---|
| 1 | DateTime | Date and time of sending the message |
| 2 | Integer | Sender id in the Timik.pl service |
| 3 | Integer | Receiver id in the Timik.pl service |

Table 6. Description of messages file.

| Column | Format | Short description |
|---|---|---|
| 1 | DateTime | Date and time of sending the message |
| 2 | Integer | Sender id in the Timik.pl service |
| 3 | Integer | Receiver id in the Timik.pl service |
| 4 | Integer | Amount of virtual currency transferred from sender to receiver |

Table 7. **Description of transactions file.**

| Column | Format | Short description |
|---|---|---|
| 1 | DateTime | Date and time of the visit in private room |
| 2 | Integer | Private room owner id in the Timik.pl service |
| 3 | Integer | Private room visitor id in the Timik.pl service |

Table 8. **Description of visits file.**

## Methods

The presented dataset covers data from dedicated viral campaigns within the virtual world, with the main focus being on spreading virtual goods. Campaigns were conducted within Timik.pl—a virtual world platform available for users in Poland, and providing chat functions and various forms of entertainment. Users are represented by graphical avatars, each having the opportunity to engage in the life of online community. Since the system launch in 2007, over 850,000 accounts have been activated. The platform was fully operational between 2007 and 2012.

All analysed campaigns were based on attempts to introduce new virtual product (graphical avatar in all cases) to the users with the use of spreading and social influence mechanics. Instead of adding new object to the system and make it available to all users, a small number of random users was selected for each campaign as a seeds and they received the products first. Then they had possibility to spread it among other users. The data collection process from the Timik.pl platform was reviewed by the Ethics Committee for Human Research at SWPS University of Social Sciences and Humanities (Wrocław Faculty, Poland). The Committee did not have any objection to the process. In order to use the platform, each user had to agree that data related to user behavior can be processed for statistical analysis and made available to external entities for analytical and research purposes after encoding, without possibility to identify any specific users. The aforementioned Ethics Committee confirmed that the dataset is in compliance with Polish law, ethical standards, and the terms of use of the Timik.pl platform.

Users communicate in the space of public graphical rooms which are associated with different themes. They can configure and furnish their own private rooms as well as use online games and different entertainment options. The basic service function relates to communication, establishing social relations, meeting new people and chatting. Avatars and decorative elements, styles, clothes and virtual products are other features. Viral content can be distributed through private messages delivered with the use of internal communication system. Analytical module enabled monitoring the spread of the content and gathering data related to information diffusion processes.

Users have access to a number of generally available functions as well as paid premium services, which provide additional possibilities. In extended packages, there are sets of avatars, clothes and special effects available. There is a store where users can buy virtual objects, and an auction system which functions as a secondary market. Free of charge and paid service elements are introduced into a system, but at different frequencies. To improve the ability to interact with other users, mechanics were used to transfer virtual goods between user account, and this was implemented for several campaigns. Virtual goods are in the form of avatars, special effects for avatars or products equivalent to real goods. Special commands implemented in the system give users the possibility to transfer virtual goods between accounts. Figure 1 illustrates the typical situation when a user who has acquired a new virtual product shows such a product to others. The interaction builds interest in the new product and it can be transmitted to other users with content spreading process observed.

The presented dataset includes data from five campaigns based on new virtual products implemented within the system. In each campaign, the new product was delivered to randomly selected users. They were able to spread it among other users, and each transmission event was registered in the system.

Designed campaigns used two mechanics to evaluate its influence on campaigns dynamics and effectiveness. Similar case was with incentives in one of campaigns. Different configurations of the campaigns were the attempt to find an optimal configuration for the best strategy for new product introduction and increasing user engagement in a new product spreading for future campaigns.

As for the campaigns, it was discovered that the following factors affected performance: content type, content transmission mechanics, content quality and incentives.

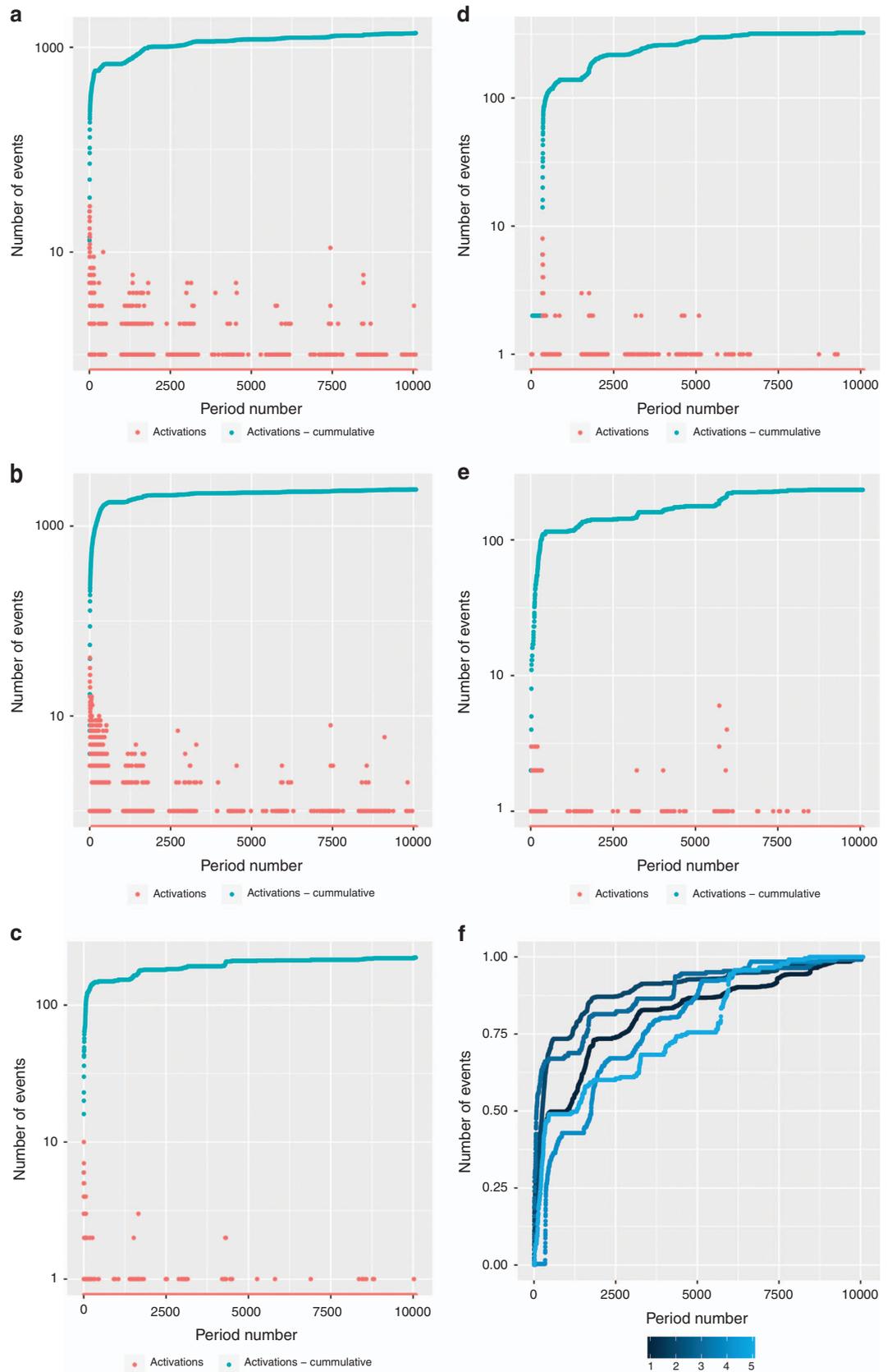

**Figure 2. Number of activations for the first seven days for campaigns C1-C5, one period is equal to one minute.** Number of activations and cumulative coverage of campaign C1 (**a**), C2 (**b**), C3 (**c**), C4 (**d**) and C5 (**e**). Normalized cumulative number of activations for campaigns C1-C5 (**f**).

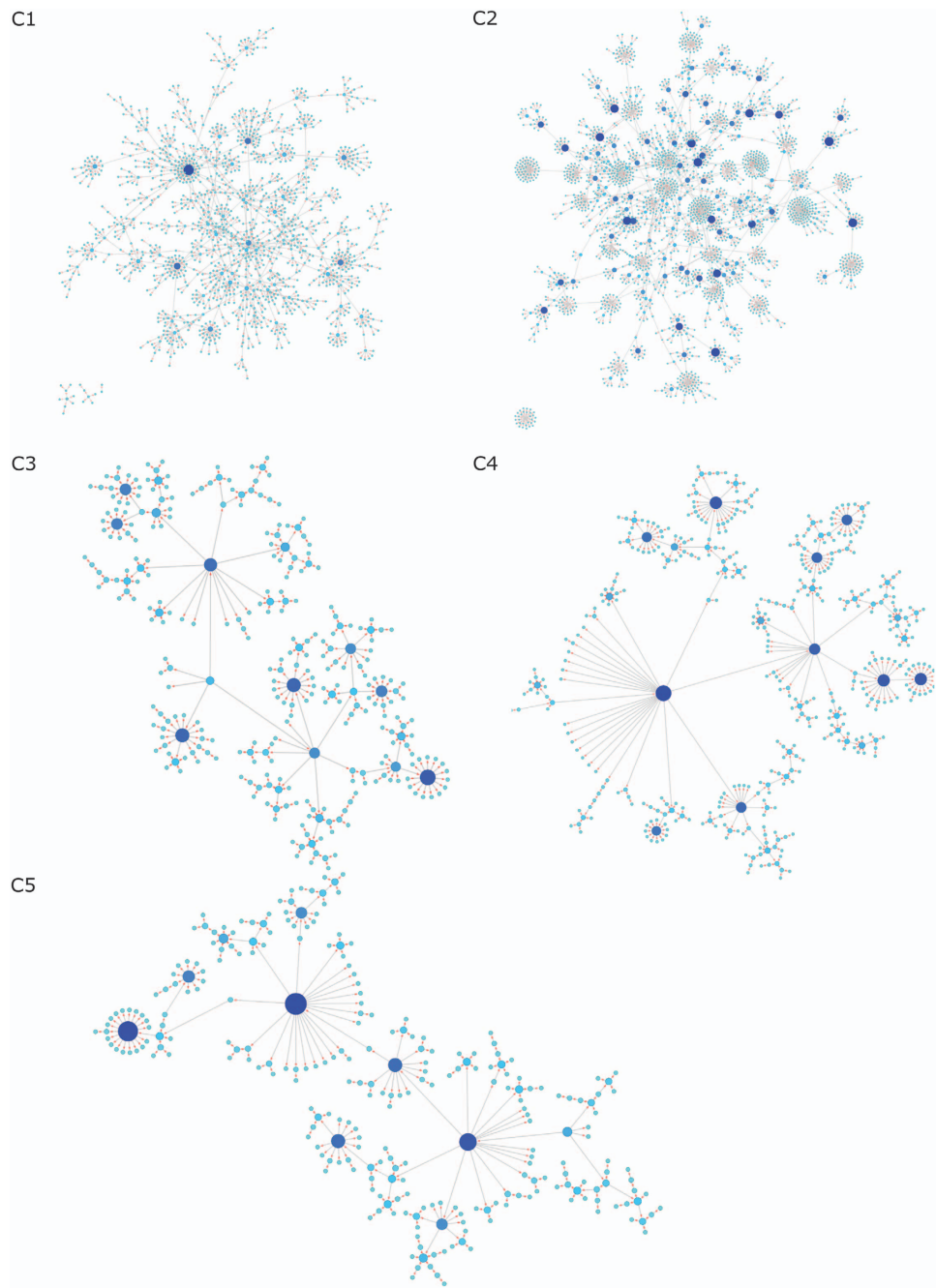

Figure 3. The visualization of campaigns C1-C5 showing how avatars diffused between users.

Content type: the virtual products designed for viral transmission were delivered in the form of thematic avatars, making users more unique within the virtual world. The avatars used for viral campaigns were different from those who were previously available within the system. The only way to get new avatars was by receiving them from the other user.

Content transmission mechanics: the asynchronous system provides opportunities to implement content sharing mechanics. Two types of mechanics were used for transmitting content between user avatars. The first possible way of content transmission used was by clicking on the recipient's avatar. It was considered as low resistance (LR) transmission with a low effort needed to transmit the content and no social relation required between sender and receiver. It was used in one campaign. The second transmission method required more effort and was based on special messages sent between users through the use of the internal messaging system. It required the existence of the receiver on a list of friends of sender and was considered as high resistance (HR) transmission. It was used in four campaigns.

Content quality: Users had access to basic avatars in the system available for free. Extended versions with higher quality were available for premium paid accounts. Differences in the quality were noticeable to users. Premium avatars were more detailed and colorful, and it was possible to distinguish among users. In four of the analysed campaigns, higher quality content (which was more similar to premium accounts) was used, and content similar to basic accounts was implemented in one of them.

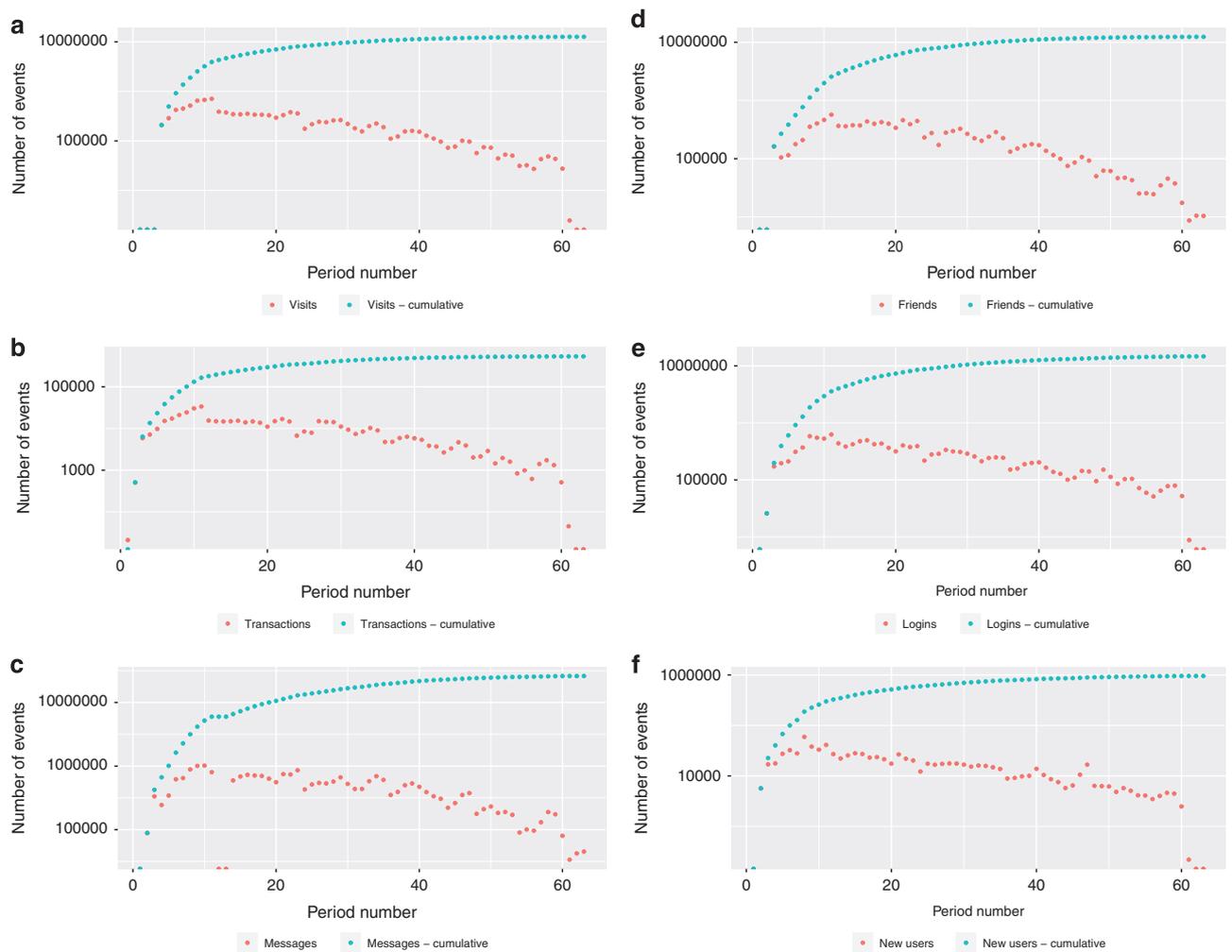

**Figure 4.** The number of different events within Timik.pl virtual world, one period is equal to one month. (**a**) Visits in virtual rooms (**b**) Transactions (**c**) Messages (**d**) Friends (**e**) Logins (**f**) New users.

Incentives: motivation to spread content to other users in viral campaigns can be based on spontaneous need of sharing content; this takes place in many online viral actions. Push and pull mechanisms are the main drivers of viral campaigns. Marketers use incentives to motivate users to share content for the purposes of increasing the dynamics of information spreading processes. In order to observe effect of incentives in one of five campaigns, incentives were used and users spreading content took parts in the contests with prizes given to the most active spreaders.

Taking previous assumptions into account, campaigns were conducted with different diffusion mechanisms and content characteristics to observe how they affected the dynamics of diffusion. The dataset included five campaigns based on the transmission of avatars within the system with specifics presented in Table 1.

Campaigns increased social relations and were making the system more interesting. It was a game where users shared knowledge about new products with friends, and established their position by possessing unique products or information. This led to the creation of a certain level of interest, emotions and demand for a product, which was not available for everyone.

### Data Records

The data is available at Harvard Dataverse (Data Citation 1) in the form of six separate files as described in Table 2. Each file contains the record of events of a given type: campaigns, friends, logins, messages, transactions and visits. The data can be used to build a multilayer network where each layer can represent a different kind of event. The structure of each data file is presented in Tables 3,4,5,6,7,8.

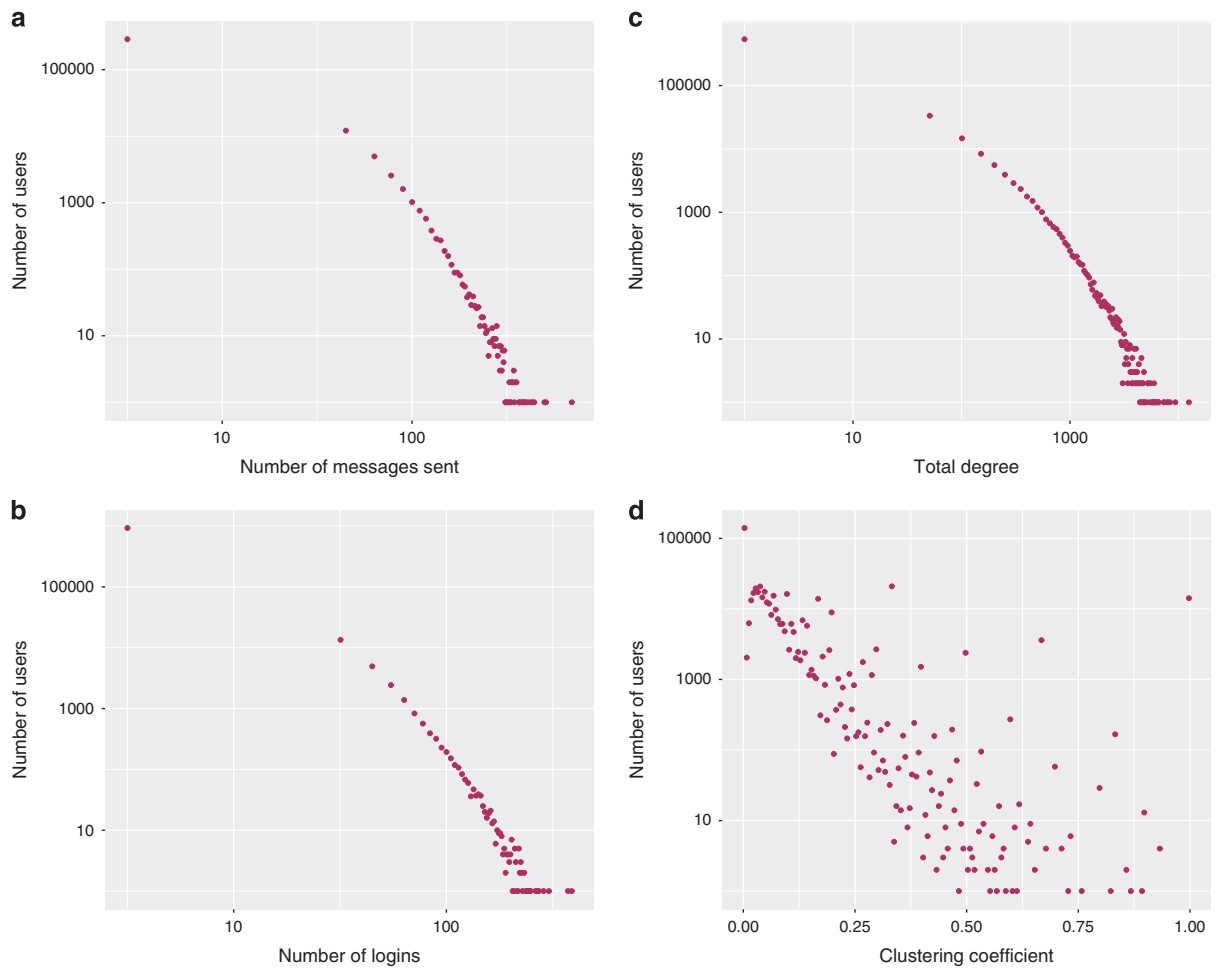

**Figure 5. The distributions of selected measures.** (**a**) Number of messages sent (**b**) Number of logins to the platform (**c**) Total degree (**d**) Clustering coefficient.

| Type | Minimum | Maximum | Average |
|---|---|---|---|
| Logins | 1 | 15,108 | 15.39 |
| Messages | 1 | 48,281 | 83.44 |
| Friends | 1 | 9,059 | 24.72 |
| Transactions | 1 | 1,931 | 6.5 |

**Table 9. The statistics of different kind of users' activities.**

### Technical Validation

The data consists of two main parts: campaigns that spread in the virtual world (C1-C5, see Table 1) and—to put the campaigns in a wider context—other events of the virtual world, such as friendships, logins to the portal, messages, virtual money transactions or visits in private rooms. The list of available data along with a brief description is presented in Table 2.

Each campaign has been analysed in terms of the number of activated users and coverage and this data is presented in Fig. 2. Figure 3 visualizes the transmissions across the network to show how the avatars spread between users.

As the campaigns are only a subset of events that occur in the virtual world, to give the researchers the opportunity to analyse how the selected events relate to all activities in the portal, the accompanying data has been also summarized as the number of events in a given period and cumulative number of events and this is presented in Fig. 4.

Moreover, the distributions of chosen measures are presented in Fig. 5, while the basic quantification of the users' activities in different layers is presented in Table 9.

### Acknowledgements

This work was partially supported by the National Science Centre, Poland, grant no. 2015/17/D/ST6/04046, 2016/21/D/ST6/02408 and 2016/21/B/HS4/01562; the European Union's Horizon 2020 research and innovation programme under the Marie SkÅ‚odowska-Curie grant agreement No. 691152 (RENOIR); and the Polish Ministry of Science and Higher Education fund for supporting internationally co-financed projects in 2016–2019 (agreement no. 3628/H2020/2016/2).

### Author Contributions

J.J. acquired the data; J.J., R.M. and P.B. preprocessed the data; J.J., R.M. and P.B. prepared the analyses. All authors contributed to the text of the manuscript.

### Additional Information

**Competing interests:** The authors declare no competing financial interests.

**How to cite this article:** Jankowski, J. *et al.* A multilayer network dataset of interaction and influence spreading in a virtual world. *Sci. Data* 4:170144 doi: 10.1038/sdata.2017.144 (2017).

**Publisher's note:** Springer Nature remains neutral with regard to jurisdictional claims in published maps and institutional affiliations.